\documentclass[10pt,aps,prb,superscriptaddress,nobibnotes,notitlepage,twocolumn]{revtex4-2}

\usepackage{pdfpages}
\usepackage{etoolbox}

\usepackage{mathptmx}       
\usepackage{mathtools}      
\usepackage{bm}             

\usepackage{graphicx}       
\graphicspath{{./Figs/}}    

\usepackage{xcolor}
\usepackage{color}          

\usepackage{nameref}        
\usepackage{hyperref}       
\hypersetup{colorlinks=true, linkcolor=blue, citecolor=blue, urlcolor=blue}

\usepackage{textcomp}

\newcommand*{\headingSectionPRL}[1]{\belowpdfbookmark{#1}{#1}{\textit{#1.---}}\ignorespaces}
\let\section\headingSectionPRL

\newcommand{\beq}[1]{\begin{equation}\label{#1}} 
\newcommand{\eeq}{\end{equation}}                
\newcommand{\eep}{\;.\end{equation}}             
\newcommand{\eec}{\;,\end{equation}}             

\renewcommand{\d}{\delta}

\newcommand{\D}{\Delta}
\newcommand{\G}{\Gamma}

\DeclareMathAlphabet{\mathcal}{OMS}{cmsy}{m}{n} 

\newcommand{\C}{\mathcal{C}}   

\newcommand{\V}{\mathcal{V}}   

\newcommand{\Ef}{\mathcal{E}}

\newcommand{\bvec}[1]{\mathbf{#1}}

\newcommand{\Vtot}{\mathcal{V}_{\mathrm{tot}}}

\makeatletter
\patchcmd{\@outputpage@head}{\@ifx{\LS@rot\@undefined}{}{\LS@rot}}{}{}{}
\makeatother

\newcommand{\NTU}{School of Electrical and Electronic Engineering, Nanyang Technological University Singapore, 50 Nanyang Avenue, 639798, Singapore}
\newcommand{\CSRC}{Beijing Computational Science Research Center, Beijing 100193, China}
\newcommand{\Beijing}{Department of Physics, Beijing Normal University, Beijing 100875, China}

\begin{document}

\title{
Hybrid Electronic-Ionic Ferroelectricity in Superlubric van der Waals Heterostructures
}

\author{Jing Huang}
\email{jing.huang@ntu.edu.sg}
\affiliation{\NTU}

\author{Jun Kang}
\affiliation{\CSRC}
\affiliation{\Beijing}

\author{Daniel Bennett}
\email{daniel.bennett@ntu.edu.sg}
\affiliation{\NTU}

\begin{abstract}
One strategy to lower the switching barrier in a sliding ferroelectric (sFE) is to insert an incommensurate spacer to reduce sliding friction, creating a superlubric sliding ferroelectric (SL-sFE). However, how polarization survives across the effectively decoupled outer layers remains an open question. We show that SL-sFEs are fundamentally different from conventional sFEs: polarization is not driven by sliding alone, but by an intricate coupling between interlayer sliding and the out-of-plane buckling of the spacer layer. This coupling results in a unique hybrid electronic-ionic polarization arising from asymmetric orbital hybridization. The interplay of these order parameters generates several distinct types of ferroelectric hysteresis, including mixed first- and second-order transitions, multi-step switching, and antiferroelectric-like behavior, establishing SL-sFEs as a distinct class of ferroelectrics.
\end{abstract}

\maketitle


\section{Introduction}
Sliding ferroelectrics (sFEs) are one of the most promising recent discoveries in the fields ferroelectricity and 2D materials. 
Originally predicted with theory and first-principles calculations~\cite{li2017binary}, they were soon after discovered experimentally~\cite{yasuda2021stacking,stern2021interfacial}, and have recently been shown to have outstanding performance in devices~\cite{yasuda2024ultrafast,bian2024developing}.
Unlike conventional ferroelectrics, the polarization in sFEs is purely electronic, originating from long-range interlayer interactions and charge transfer (about 1\% of an electron~\cite{bennett2022electrically,bennett2022theory}). 
Because of this unique mechanism, the polarization in sFEs gives rise to a wide variety of novel physics in 2D materials, for example moiré polar domains~\cite{stern2021interfacial,yasuda2021stacking,wang2022interfacial,weston2022interfacial,molino2023ferroelectric,van2024engineering,wang2025moire}, which have a nontrivial response to electric fields field~\cite{bennett2022electrically,bennett2022theory,ko2023operando,bennett2024asymmetric,ramos2025flat}, nontrivial polar topology~\cite{bennett2023polar,bennett2023theory,jankowski2024polarization,vu2024imaging}, as well as multiferroic order~\cite{sivadas2018stacking,poudel2023creating,bennett2024stacking}.
Notably, the electronic nature of sFEs is also what makes them remarkable for applications: 
because the switching relies on long-range interactions and sliding rather than breaking and forming bonds, the ferroelectricity is intrinsically fatigue-free~\cite{yasuda2024ultrafast,bian2024developing}.
This unconventional nature of polarization and ferroelectricity in sFEs also introduces new challenges for their design and optimization in devices.

One of the primary challenges for the optimization of sFEs in devices is the lowering of the coercive field required for ferroelectric switching.
This field is determined by the spontaneous polarization, which is intrinsically weak due to its electronic and long-range nature, and the energy barrier for switching, which is typically large for global, rigid sliding. 
For ideal rigid sliding in hexagonal boron nitride (hBN) and transition metal dichalcogenides (TMDs) such as MoS$_2$, WSe$_2$, etc., the theoretical coercive field is of order 2~V/\AA{}~\cite{bennett2022electrically,bennett2022theory}, which is approximately 100 times larger than experimentally observed values~\cite{yasuda2021stacking,ko2023operando}.
It is far more energetically efficient to switch polarization through the motion of domain wall stackings, partial dislocations separating AB and BA stackings (polarization up and down), that sweep through the system like solitons in response to an electric field~\cite{yasuda2021stacking,wang2022interfacial,ko2023operando,wang2025polarization,ke2025superlubric}.
These domain walls are topologically protected and cannot nucleate spontaneously~\cite{engelke2023non};
they must be included in a sample manually, typically by imposing a marginal twist angle (even in ``untwisted'' systems)~\cite{yasuda2021stacking,wang2022interfacial,ko2023operando,yasuda2024ultrafast}.
One promising alternative strategy to lower the sliding energy barrier is the recent proposal of the superlubric sliding ferroelectric (SL-sFE)~\cite{yang2025superlubric}, in which a spacer layer with a different lattice constant is inserted into the middle of a sFE to create a superlubric interface.

By sliding over the spacer layer without friction, this dramatically lowers the energy barrier~\cite{yang2025superlubric}.
While this sounds promising, there is one conceptual issue: inserting a spacer approximately doubles the distance between the outer layers.
In the absence of a spacer, the outer layers would effectively be decoupled, and their relative stacking alone could not generate a spontaneous polarization.
Therefore, SL-sFEs cannot be thought of simply as a generalization of sFEs with a reduced energy barrier.
The physical mechanism for polarization and ferroelectricity must be fundamentally different from those of sFEs.
Because the outer layers would otherwise be decoupled, the spacer itself must be playing an active role in not only lowering the energy barriers, but also mediating polarization and ferroelectricity.

\begin{figure*}[htbp]
\centering
\includegraphics[width=1.0\linewidth]{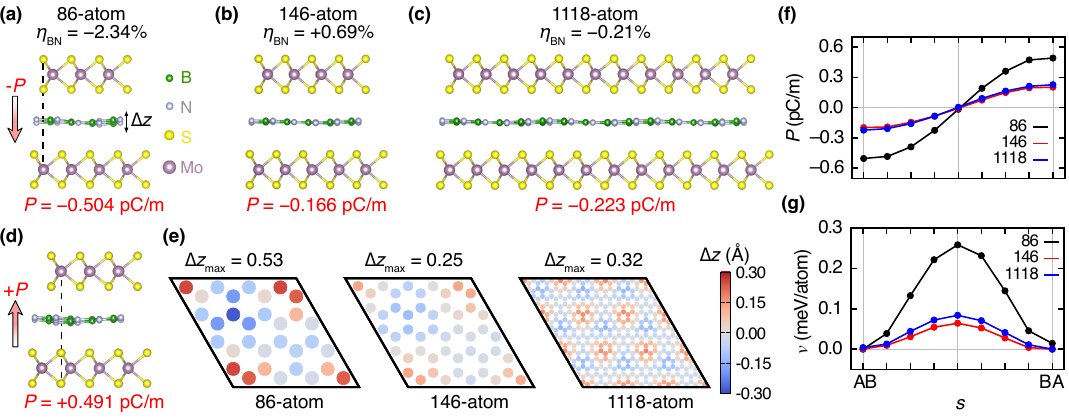}
\vspace{-1.5em}
\caption{%
{\bf (a)}--{\bf (c)} MoS$_2$/hBN/MoS$_2$ heterostructures for {\bf (a)} 86-atom, {\bf (b)} 146-atom, and {\bf (c)} 1118-atom commensurate supercells, where $\D z$ represents the out-of-plane buckling of the buckled spacer.
The biaxial strain $\eta_{\rm BN}$ imposed on the hBN spacer in each supercell is shown above, and the out-of-plane polarization $P$ is shown below in red.
{\bf (d)} The 86-atom supercell after sliding of the outer layers, which reverses the direction of the polarization, indicated by the red arrow.
{\bf (e)} Out-of-plane buckling of the atoms in the hBN spacer layer for the three supercells, where $\D z_{\mathrm{max}}$ is the maximum displacement.
{\bf (f)} Polarization $P$, and {\bf (g)} relative change in energy $\nu$ as a function of the NEB parameter ($s$) connecting the polarization up (BA) and down (AB) states for the 86-atom (black), 146-atom (red) and 1118-atom (blue) supercells.
}
\label{Fig1}
\end{figure*}

In this Letter, we show using theory and first-principles calculations that the physical properties of superlubric sliding ferroelectrics are fundamentally different from those of conventional sFEs.
Contrary to the original proposal~\cite{yang2025superlubric}, polarization does not originate from sliding alone, but from an intricate interplay between interlayer sliding of the outer layers and out-of-plane buckling of the spacer layer.
The polarization of unique hybrid electronic-ionic origin:
rather than being purely electronic as in normal sFEs, it arises from an asymmetric orbital hybridization between the spacer and the outer layers, which is controlled by the combination of sliding and buckling.
The interplay between sliding and buckling order parameters leads to a rich parameter space with multiple distinct types of ferroelectric hysteresis, including mixed first- and second-order transitions, and multi-step switching, making them a novel class of ferroelectrics distinct from conventional sFEs.

\section{Results}
First-principles simulations were performed (see Supplementary Material, SM), on MoS$_2$/hBN/MoS$_2$ (MBM), a prototypical SL-sFE, see Fig.~\ref{Fig1}.
While the original proposal suggested that introducing a spacer could lower the sliding barrier by an order of magnitude~\cite{yang2025superlubric}, we find that the physical properties of SL-sFEs are highly sensitive to the choice of commensurate cell;
constructing a periodic SL-sFE for simulations inherently requires imposing a lattice mismatch strain between the outer and spacer layers (MoS$_2$ and hBN in this case).
Figs.~\ref{Fig1} (a)-(c) show three MBM supercells of increasing size, each with a different biaxial strain $\eta_{\rm BN}$ on the hBN spacer layer, and out-of-plane polarization $P$.
Using the nudged elastic band (NEB) method~\cite{henkelman2000climbing} to connect the polarization up and down states through sliding, we find that as $\eta_{\rm BN}$ increases, both the polarization and stacking energy~\cite{kaxiras1993free} are significantly reduced, see Figs.~\ref{Fig1} (f) and (g).
This shows that there is a trade-off between lowering the energy barrier and sustaining the out-of-plane polarization, which was previously overlooked as only a single commensurate supercell with a relatively large lattice mismatch was considered (the 86-atom supercell considered here).

\begin{figure*}[ht]
\centering
\includegraphics{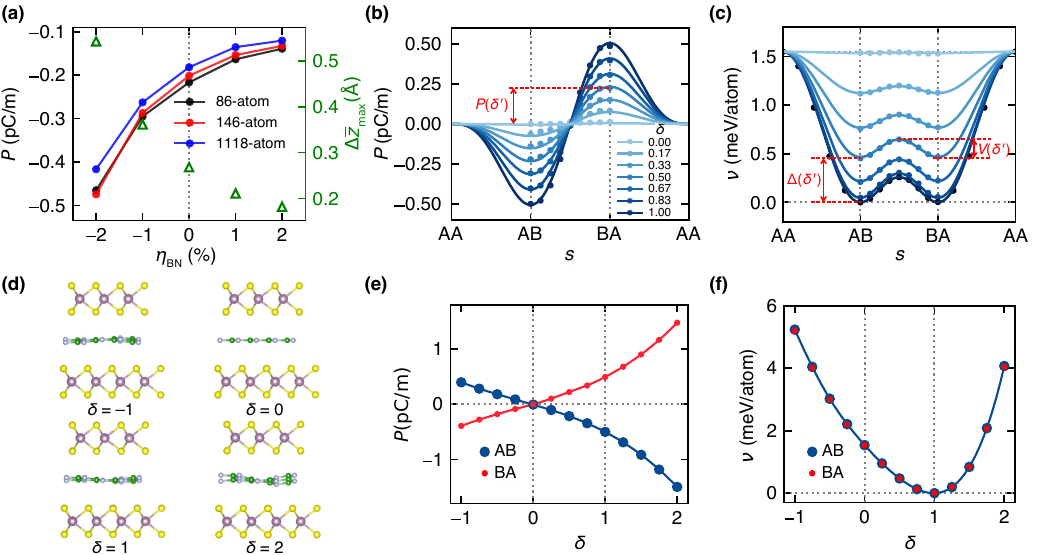} 
\vspace{-0.5em}
\caption{%
\textbf{(a)} Out-of-plane polarization $P$ as a function of biaxial strain imposed on the hBN spacer layer for 86-atom (black), 146-atom (red) and 1118-atom (blue) MBM supercells.
The average out-of-plane atomic displacement $\D \bar{z}$ of the spacer is shown in green.
\textbf{(b), (c)} Polarization $P$ and relative energy $\nu$ along the NEB path ($s$) for different fixed values of $\d$.
The data points are results obtained from first-principles calculations, and the solid lines represent fitting to the analytic model described by Eq.~\eqref{eq:model}.
The model parameters $V$, $P$ and $\Delta$ from Eq.~\eqref{eq:model} are highlighted in red for $\d'=0.5$.
\textbf{(d)} Side views of the AB stacked MBM structures in \textbf{(b)} and \textbf{(c)} at $\d=-1,0,1,2$.
\textbf{(e), (f)} $P$ and $\nu$ as a function of buckling for the AB (blue) and BA (red) stackings.
}
\label{Fig2}
\end{figure*}

The apparent dependence on the supercell size actually originates from the lattice mismatch imposed in the SL-sFE, which results in an out-of-plane buckling of the spacer layer in order to accommodate the induced strain (Fig.~\ref{Fig1} (e)).
To verify that the lattice mismatch and not the supercell size that affects physical properties, we calculated the polarization as a function of biaxial strain applied to the hBN spacer layer in all three supercells, see Fig.~\ref{Fig2} (a).
When the hBN spacer is unstrained, the polarization is approximately 0.2~{pC/m} for all cases: about 30\text{--}40\% of that of parallel-stacked bilayer MoS$_2$.
Despite their different lattice mismatches, all three supercells collapse onto a single curve, illustrating that the strain on the spacer, rather than supercell size, influences the polar properties of SL-sFEs.
The reason for this is that the strain leads to buckling of the spacer layer.
We also note that the outer MoS$_2$ layers have little effect on the polarization, as all three cells, which have different strain on MoS$_2$, yield practically the same polarization.

To isolate the effect of strain $\eta_{\rm BN}$ on the polarization and energetics, we introduce a generalized coordinate $\d$ which parametrizes the buckling continuously from a perfectly flat layer ($\d = 0$) to a buckled layer in the equilibrium structure ($\d = 1$).
Structures in between $\d=0,1$ are obtained by linearly interpolation, and structures outside of this range are obtained through extrapolation (Fig.~\ref{Fig2} (d)).
NEB calculations were performed to map the transition pathway between the polarization-up and polarization-down states with sliding, for a series of fixed values of $\delta$ ranging from 0 to 1.

The resulting out-of-plane polarization and stacking energies along the NEB paths are shown in Figs.~\ref{Fig2} (b) and (c), respectively.
The functional form of the polarization and stacking energy are the same as those of conventional sFEs~\cite{carr2018relaxation,bennett2022electrically,bennett2022theory}, but their magnitudes are controlled by the buckling of the spacer.
As the spacer is flattened from its equilibrium structure ($\d = 1$) to a perfectly flat layer ($\d = 0$), the amplitude of the polarization curve smoothly changes to zero.
Furthermore, the buckling has two distinct effects on the stacking energy landscape:
First, the local barrier separating the two polar states vanishes as $\d\to 0$.
Second, there is a uniform shift of the entire stacking energy landscape as the buckling decreases, corresponding to the elastic energy $\D(\d)$ required to suppress the spontaneous out-of-plane buckling of the spacer.

\begin{figure*}[htbp]
\centering
\includegraphics{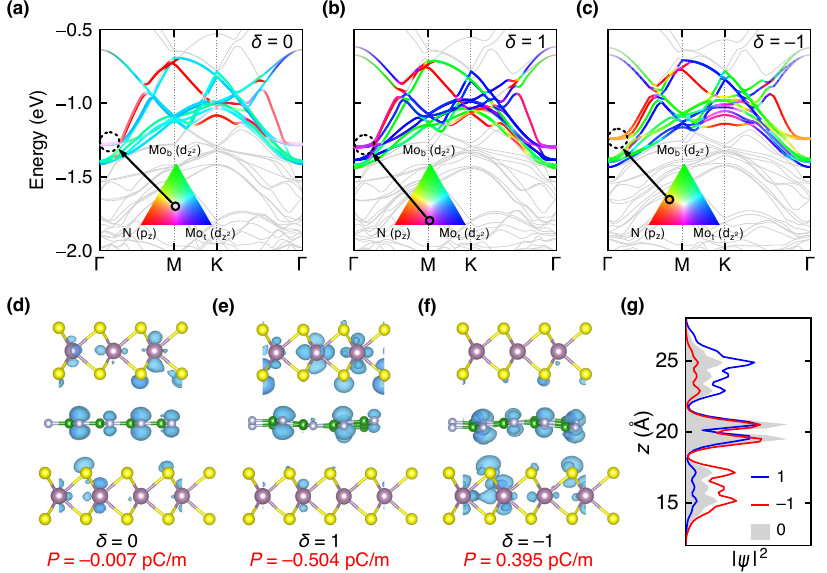}
\vspace{-1em}
\caption{%
\textbf{(a)}--\textbf{(b)} Electronic band structure of an 86-atom MBM supercell for {\bf a} $\delta=0$, {\bf (b)} $\delta=1$ and {\bf (c)} $\delta=-1$.
The orbital projection of the low-energy valence bands are shown, projecting onto the N $p_z$ states (red), and Mo $3d_{z^{2}}$ states of the bottom (green) and top (blue) layers.
The highlighted states at $\G$ show hybridization between {\bf (a)} all three layers (white), {\bf (b)} the spacer and top layer (pink), and {\bf (c)} the spacer and bottom layer (orange).
\textbf{(d)--(f)} Wavefunction of the state at $\G$ highlighted in the band structure.
\textbf{(g)} Magnitude squared of the same wavefunctions in the out-of-plane direction, averaged in-plane, for $\d=0$ (gray region), $\d=1$ (blue line) and $\d=-1$ (red line).
}
\label{Fig3}
\end{figure*}

The polarization and stacking energy are directly related to $\d$ even outside the range $[0,1]$, for example if the buckling is enhanced ($\d = 2$), or reversed ($\d=-1$), see Fig.~\ref{Fig2} (d).
The polarization and stacking energy as a function of $\d$ are shown in Figs.~\ref{Fig2} (e) and (f) for fixed AB (blue) and BA (stackings).
Notably, this highlights that polarization switching can occur \textit{without sliding}, solely from the buckling of the spacer layer, unlike conventional sFEs.
To further verify that buckling of the spacer is responsible for the emergence of polarization, we performed calculations for a MBM structure consisting of an anti-parallel 2H-MoS$_2$ bilayer (see SM).
Although the bilayer is centrosymmetric in isolation,
inserting a spacer, which buckles in the relaxed structure, yields a nonzero polarization.

In order to understand the physical origin of polarization in SL-sFEs, and the nature of its dependence on sliding and buckling, the electronic structure of MBM was analyzed in detail.
Figs.~\ref{Fig3} (a)-(c) show the low-energy valence bands of 86-atom MBM for a perfectly flat spacer ($\d=0$), a relaxed spacer ($\d=1$), and a spacer with the buckling reversed ($\d=-1$).
When the spacer is perfectly flat ($\d = 0$), the N $p_z$ states hybridize evenly with the Mo $3d^{2}_{z}$ states of the top and bottom layers (Fig.~\ref{Fig3} (d)).
In the relaxed buckled state ($\d = 1$), symmetry is broken, and the N $p_z$ states hybridize almost exclusively with the Mo $3d^{2}_{z}$ states of the top layer (Fig.~\ref{Fig3} (e)).
Reversing the orientation of the buckling ($\d = -1$), the N $p_z$ states hybridize with the Mo $3d^{2}_{z}$ states of the bottom layer (Fig.~\ref{Fig3} (f)).
This asymmetric distribution can be seen from the magnitude squared of the low-energy $\G$-wavefunction, see Fig.~\ref{Fig3} (g).
For $\d=0$, the polarization is nearly zero, and changing the sign $\d=+1\to -1$ reverses the sign of the polarization.
This shows that the nature of polarization and ferroelectricity in SL-sFEs is of hybrid electronic-ionic origin, with structural distortions modulating the asymmetry of interlayer orbital hybridization.

Because the sliding barrier and polarization are sensitive to both sliding and buckling, the switching mechanisms in SL-sFEs are unconventional in comparison to regular sFEs.
Typically, an external electric field $\Ef$ tilts the stacking energy landscape, and switching occurs when the electrostatic energy $-\Ef\cdot p$, where $p$ is the out-of-plane dipole moment, overcomes the energy barrier.
In SL-sFEs however, the energy landscape evolves in a more complex way due to the interplay between sliding and buckling.
To capture the competition between the sliding energy barrier, electrostatic energy and elastic penalty of deforming the spacer, we propose a continuum model inspired by those used to describe moiré materials~\cite{bennett2022electrically,bennett2022theory,ramos2025flat}.
The total energy $\Vtot(\bvec{s}, \d, \Ef)$ as a function of stacking ($\bvec{s}$) and buckling ($\d$) order parameters is given by 
\beq{eq:model}
\Vtot(\bvec{s}, \d, \Ef) = \V(\d)\phi^{\rm e}(\bvec{s}) - \Ef p(\d)\phi^{\rm o}(\bvec{s}) + \D(\d)
\eep
$\phi^{\rm e}$ and $\phi^{\rm o}$ are $\C_3$-symmetric even and odd basis functions, which depend only on stacking~\cite{bennett2022electrically,bennett2022theory,ramos2025flat}.
The coefficients $V(\d)$ and $p(\d)$ capture the magnitude of the sliding barrier and out-of-plane dipole moment, and $\D(\d)$ is the energy cost to deform the spacer from its equilibrium state ($\d=1$).
This minimal model yields a remarkably good fit to our first-principles calculations for MBM (Figs.~\ref{Fig2} (b)-(c)), allowing us to extract the functional forms of $V(\d)$, $p(\d)$ and $\D(\d)$ (see SM for additional details).

Depending on the material parameters, the analytic model described by Eq.~\eqref{eq:model} yields a rich variety of ferroelectric behavior.
The overall response of the system is governed by the relative values of three critical electric fields: (i) the coercive field required to induce interlayer sliding, $\Ef_{s}$; (ii) the required field for maximum spacer buckling ($\d_{\rm m}$), $\Ef^{\rm m}_{\d}$; and (iii) the field required to perfectly flatten the spacer ($\d = 0$), $\Ef^{0}_{\d}$.
We identify four distinct types of hysteresis loops, labeled I--IV, depending on the order in which the critical fields occur.

\begin{figure*}[htbp]
\centering
\includegraphics{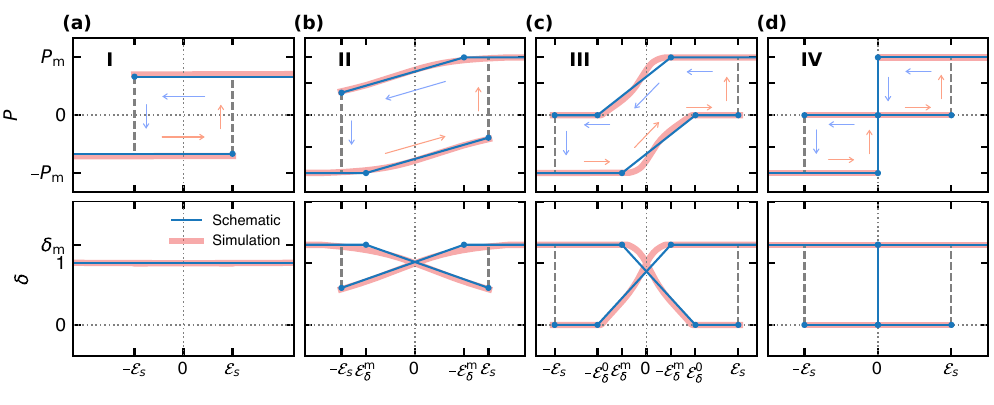}
\vspace{-1em}
\caption{%
\textbf{(a)}--\textbf{(d)} ferroelectric hysteresis loop $P(\Ef)$ (top) and corresponding buckling order parameter $\d(\Ef)$ for the four types I--IV in the analytic model Eq.~\eqref{eq:model}.
The three critical electric fields $\Ef_{s}$, $\Ef^{\rm m}_{\d}$ and $\Ef^{0}_{\d}$ are shown on the horizontal axes, where $P_\mathrm{m}$ is the polarization at $\Ef^{\rm m}_{\d}$.
The blue lines represent a schematic description of hysteresis, 
while the thick orange lines denote the results obtained from the model calculations, fitting Eq.~\eqref{eq:model} to the first-principles calculations of the 86-atom MBM supercell.
}
\label{Fig4}
\end{figure*}

Fig.~\ref{Fig4} shows the four hysteresis loops I--IV as well as the corresponding buckling order parameter $\d(\Ef)$ below.
The blue lines represent a sketch of the phase transitions, whereas the orange lines are numerical results obtained from minimizing the model fit to first-principles calculations of MBM.
For type I (Fig.~\ref{Fig4} (a)), $\Ef_{s}$ (sliding) occurs first, and there is insufficient energy to modify the buckling before the sliding is initiated.
The spacer is effectively stiff, and the system behaves like a conventional sFE, with a first-order hysteresis loop~\cite{yasuda2021stacking,yasuda2024ultrafast}.
For type II (Fig.~\ref{Fig4} (b)), $\Ef^{\rm m}_{\d}$ occurs first, and the spacer begins to un-buckle before sliding occurs. 
Sliding is accompanied by a discontinuous jump of the buckling, which results in a hysteresis loop which is of mixed first- and second-order.
For type III in Fig.~\ref{Fig4} (c), $\Ef^{\rm 0}_{\d}$ occurs before $\Ef_{s}$, and the spacer becomes perfectly flat before sliding occurs.
This resembles type II, but with a plateau at $P=0$ and $\d=0$ before switching occurs.
Finally, type IV in Fig.~\ref{Fig4} (d) represents a critical case when $\Ef^{\rm m}_{\d}=\Ef^{0}_{\d}=0$, which corresponds to the softening of the buckling mode of the spacer, resulting in an antiferroelectric-like double hysteresis loop.
Fitting the model to first-principles calculations, we find that MBM is of type II, illustrated by the orange curves in Fig.~\ref{Fig4} (b).
For illustrative purposes, other three types were generated by manually scaling the buckling energy $\D(\d)$.

\section{Discussion}
Superlubric sliding ferroelectrics were originally proposed as a strategy to lower the energy barrier in sliding ferroelectrics~\cite{yang2025superlubric}.
Although it was previously thought that this reduction of the barrier could be achieved without sacrificing the spontaneous polarization, we have demonstrated that this conclusion was an artifact of the specific commensurate supercell chosen, and the artificial strain imposed on the layers.
In reality, the physical mechanisms in SL-sFEs are much richer, and rather than simply being a low-friction generalization of conventional sFEs, they form a distinct class of ferroelectrics. 

The polarization is not mediated by sliding alone, but by an intricate coupling between sliding and buckling of the spacer layer.
This buckling can induce polarization even in systems that would be centrosymmetric in isolation (e.g.~2H-MoS$_2$), and reversing the buckling can switch the polarization without sliding.
The polarization has a unique hybrid electronic-ionic origin: the combination of sliding and buckling (ionic) breaks interfacial symmetry, which drives an asymmetric interlayer hybridization of the orbitals between the spacer layer and the outer layers (electronic).
The intricate coupling between sliding and buckling order parameters gives rise to a rich variety of ferroelectric behavior, including mixed first- and second-order transitions, and multi-step switching.

While SL-sFEs can indeed be used to lower sliding energy barriers, we find that this generally comes at the cost of reducing the spontaneous polarization.
By exploring the vast material space of 2D vdW heterostructures, it may be possible to design a SL-sFE with optimal conditions, minimizing the energy barrier with a minimal reduction of polarization.
The development of SL-sFEs would be a useful step towards realizing intrinsic ferroelectric switching in 2D materials, without the need for extrinsic effects such as domain walls, which would be highly desirable for reliability and scalability in nanodevices.
Finally, beyond technological potential, we propose that SL-sFEs may be used as a rich material platform to observe new fundamental phenomena, in particular the novel ferroelectric phases identified in this work.

\section{Acknowledgments}
D.B.~and J.H.~acknowledge support from the NTU Startup Grant (Award Number 025661-00003).
J.K.~is supported by the National Natural Science Foundation of China (12393831). 
We acknowledge computational support from the National Supercomputing Centre Singapore, High-Performance Computing Centre at Nanyang Technological University Singapore, and National Supercomputer Center in Tianjin.


%

\clearpage



\section*{Supplementary material}

# Hybrid Electronic-Ionic Ferroelectricity in Superlubric van der Waals Heterostructures

Jing Huang,[1,*] Jun Kang,[2,3] and Daniel Bennett[1,†]

[1]*School of Electrical and Electronic Engineering, Nanyang Technological University Singapore, 50 Nanyang Avenue, 639798, Singapore*
[2]*Beijing Computational Science Research Center, Beijing 100193, China*
[3]*Department of Physics, Beijing Normal University, Beijing 100875, China*

## I. FIRST-PRINCIPLES CALCULATIONS

First-principles were performed using density functional theory (DFT) with projector augmented wave (PAW) pseudopotentials, using the Vienna Ab-initio Simulation Package (VASP) [1]. A plane-wave basis with an energy cutoff of 500 eV was used in all calculations. The PBE generalized gradient approximation (GGA) functional [2] to describe the exchange-correlation energy, and van der Waals interactions were treated using the DFT-D2 correction [3]. Self-consistent field calculations were performed, using an energy tolerance of $10^{-6}$ eV. Geometry relaxations were performed, relaxing the positions of the atoms until all of the residual forces were less than 0.01 eV Å$^{-1}$. A vacuum spacing of 25 Å was included in the out-of-plane direction, and a dipole correction was included to eliminate spurious electrostatic interactions between periodic images [4].

Commensurate $MoS_2$/hBN/$MoS_2$ (MBM) supercells were constructed as follows: $3 \times 3 \times 1$ $MoS_2$ plus $4 \times 4 \times 1$ hBN for the 86-atom supercell, $4 \times 4 \times 1$ $MoS_2$ plus $5 \times 5 \times 1$ hBN for the 146-atom supercell, and $11 \times 11 \times 1$ $MoS_2$ plus $14 \times 14 \times 1$ for the 1118-atom supercell. Brillouin-zone integrations were performed using Monkhorst–Pack $k$-point meshes of $3 \times 3 \times 1$ for the 86-atom supercell, $2 \times 2 \times 1$ for the 146-atom supercell, and $\Gamma$-point-only sampling for the 1118-atom supercell. The switching energy barriers were evaluated using the climbing-image nudged elastic band (CI-NEB) method [5]. The electric polarization was obtained by calculating the Berry phase [6].

## II. ROLE OF THE SPACER

To investigate whether the inner BN spacer in the MBM system contributes to the polarization, or merely serves to reduce friction, we performed a series of 180° rotations of either the BN layer or one of the $MoS_2$ layers, as shown in Fig. S1. Fig. S1 (a) shows MBM with parallel $MoS_2$ in the AB stacking, and Fig. S1 (b) shows MBM with the same stacking of $MoS_2$, but with the hBN layer rotated by 180°. Although the alignment of the outer $MoS_2$ layers is fixed, rotating the hBN spacer switches the polarization, without any interlayer sliding. Fig. S1 (c) shows MBM with anti-parallel $MoS_2$, which has a finite polarization. In this case, the polarization cannot be switched by interlayer sliding. In the absence of the spacer, the $MoS_2$ layers would be centrosymmetric, illustrating that the spacer plays a key role in mediating polarization in superlubric sliding ferroelectrics

To investigate whether the polarization in MBM could arise from buckling of the spacer alone, we decomposed MBM into two isolated subsystems: the $MoS_2$ bilayer, and the hBN spacer, keeping the atomic structures fixed to those of the full system, see Fig. S2. In both cases, the polarization is nearly zero. This clearly illustrates that the polarization does not originate from either interlayer sliding of the outer layers or out-of-plane buckling of the spacer in isolation, but from the combined effects, as well as electronic interactions between the outer layers and the spacer.

* jing.huang@ntu.edu.sg
† daniel.bennett@ntu.edu.sg

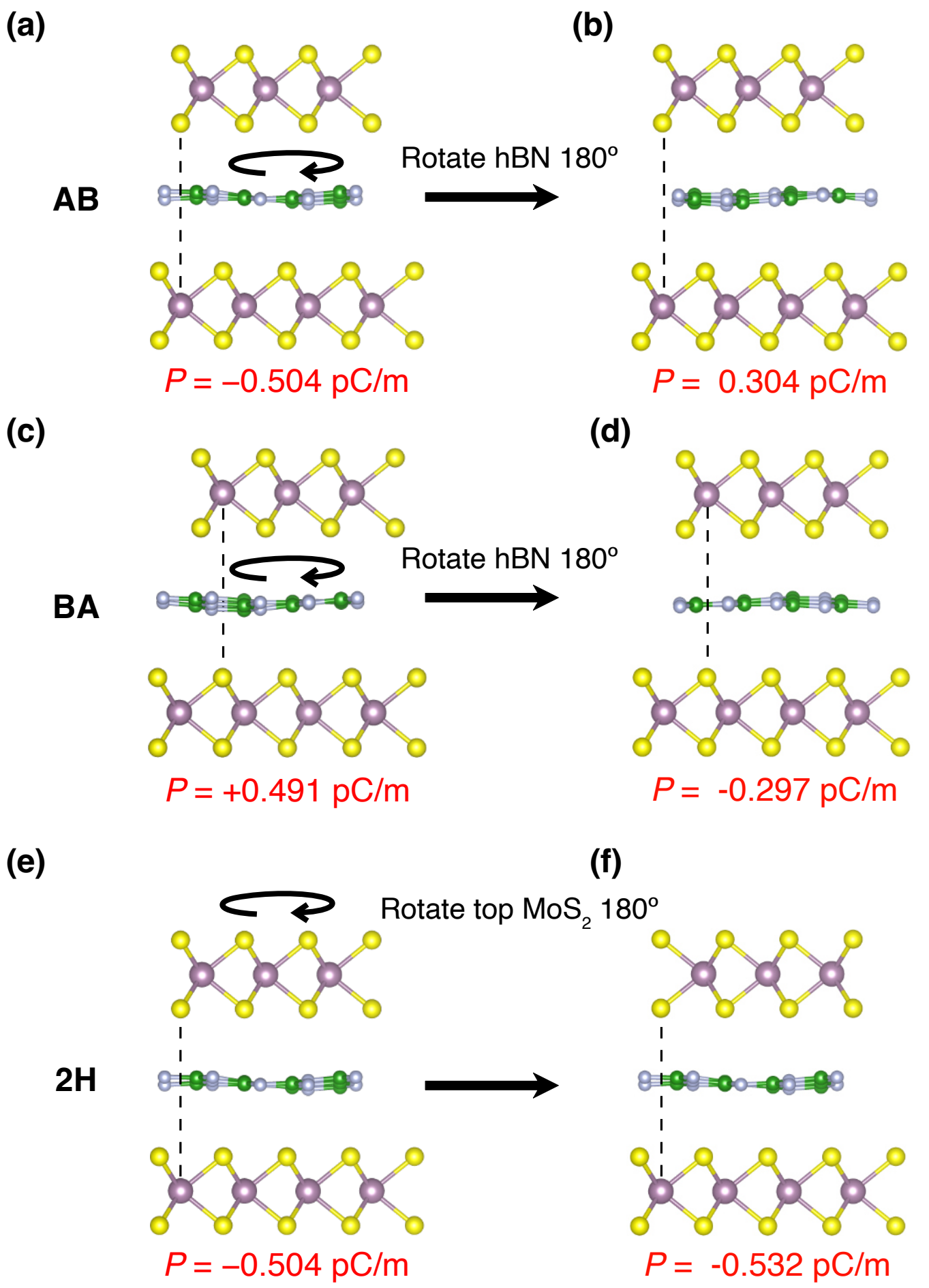


FIG. S1. **Role of the relative alignment of the layers.** **a** 86-atom MBM with parallel $MoS_2$ layers in the AB stacking. **b** The same structure, but with the hBN spacer rotated by 180°, which reverses the polarization. **c** 86-atom MBM with an anti-parallel arrangement of the $MoS_2$ layers. The orientation of the spacer with respect to the bottom layer is the same as in (a).

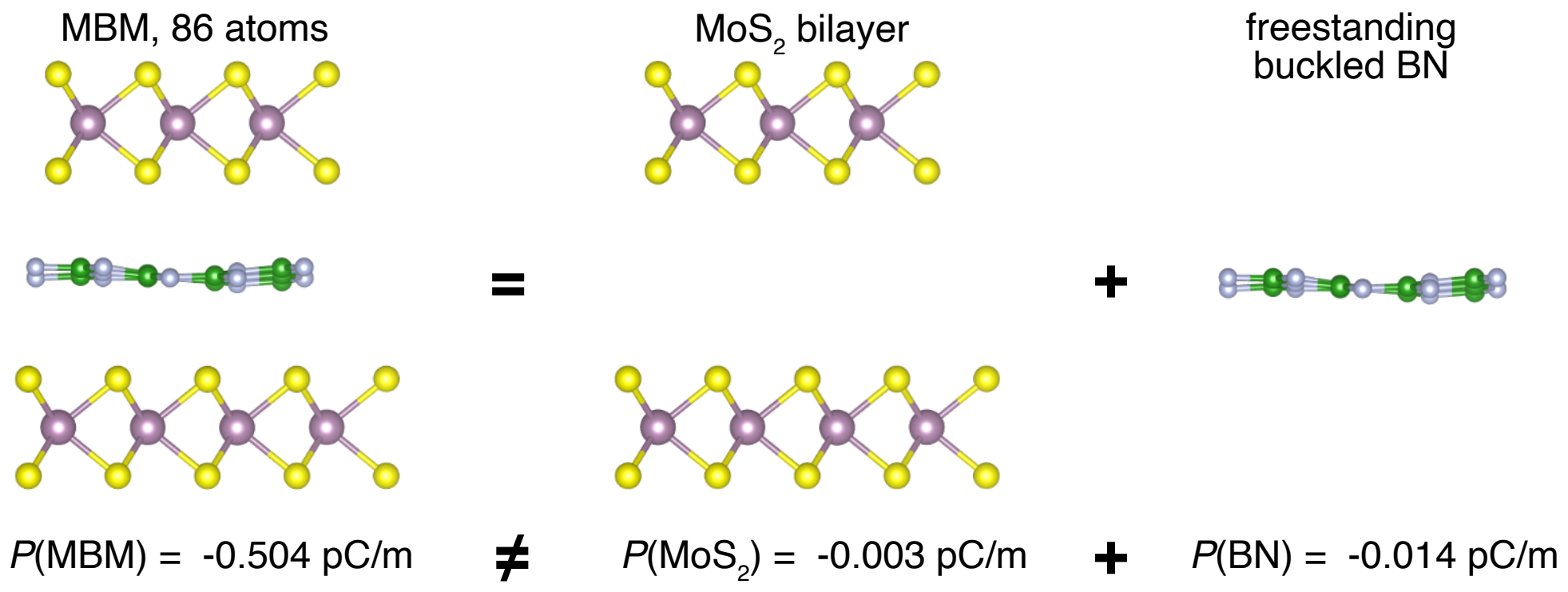


FIG. S2. **Decomposition of the MBM system.** The left shows the 86-atom MBM supercell, and the right shows its decomposition into the outer $MoS_2$ layers and the hBN spacer in isolation. The individual sub-systems have negligible polarization.

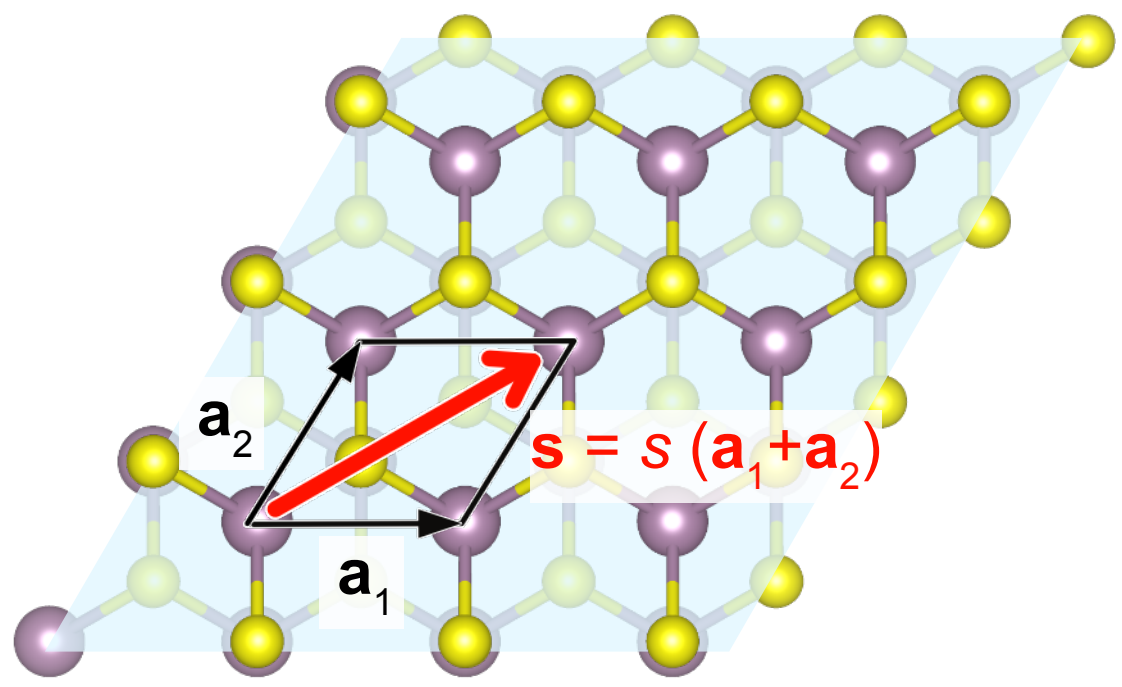


FIG. S3. **The high-symmetry path in a MBM supercell.** The semi-transparent blue layer corresponds to the BN spacer, where the red arrow indicates the high-symmetry sliding path **s** along the $MoS_2$ armchair direction, the black arrows $\mathbf{a}_1, \mathbf{a}_2$ are the lattice vectors of the $MoS_2$ outer layers.

## III. THEORETICAL MODEL

We propose a Landau-like model of SL-sFEs, inspired by the elastic continuum models used to describe moiré materials [7–9]. The energy $\mathcal{V}$ is described by two order parameters: **s**, the relative stacking of the outer layers (corresponding in this work to the high-symmetry sliding path of the $MoS_2$ outer layers along the armchair direction), and $\delta$, the buckling of the spacer layer. The total energy is comprised of a balance between three contributions:

$$\mathcal{V}_{\text{tot}}(\mathbf{s},\delta,\mathcal{E}) = \mathcal{V}_{\text{stack}}(\mathbf{s},\delta) + \mathcal{V}_{\text{elec}}(\mathbf{s},\delta,\mathcal{E}) + \mathcal{V}_{\delta}(\delta)\ . \tag{1}$$

Here, we confine the stacking vector to the $MoS_2$ unit cell diagonal, $\mathbf{s} = s(\mathbf{a}_1 + \mathbf{a}_2)$, see Fig. S3 The unit cell diagonal contains all of the high symmetry stackings: AA ($s = 0$), AB ($s = \frac{1}{3}$), BA ($s = \frac{2}{3}$) and domain wall (DW, ($s = \frac{1}{2}$)), and is sufficient to obtain a good description of the system for all **s** [7, 8]. The first term is the stacking energy, or cohesive energy between the layers, which depends primarily on **s**, but it also influenced by buckling $\delta$. Assuming a separation of variables, we write the stacking energy as

$$\mathcal{V}_{\text{stack}}(\mathbf{s},\delta) = \mathcal{V}(\delta)\sum_n \Phi_n^{\text{e}}(\mathbf{s})\ . \tag{2}$$

$\Phi_n^{\text{e}}(\mathbf{s})$ are $\mathcal{C}_3$-symmetric basis functions which are even in stacking: $\Phi_n^{\text{e}}(\mathbf{s}) = \sum_{\mathbf{g}_i^n} \cos(\mathbf{g}_i^n \cdot \mathbf{s})$, where the sum runs over reciprocal lattice vectors in the $n^{\text{th}}$ star, $\{\mathbf{g}_i^n\}$ [7–9]. Similarly, the electrostatic energy can be written as

$$\mathcal{V}_{\text{elec}}(\mathbf{s},\delta,\mathcal{E}) = -\mathcal{E}p(\mathbf{s},\delta) = -\mathcal{E}p(\delta)\sum_n \Phi_n^{\text{o}}(\mathbf{s})\ , \tag{3}$$

where $\Phi_n^{\text{o}}(\mathbf{s}) = \sum_{\mathbf{g}_i^n} \sin(\mathbf{g}_i^n \cdot \mathbf{s})$ are $\mathcal{C}_3$-symmetric odd basis functions. For simplicity we use only the first star in the expansions of $\mathcal{V}_{\text{stack}}$ and $p$:

$$\begin{aligned} \mathcal{V}_{\text{stack}}(\mathbf{s},\delta) &= \mathcal{V}(\delta)\Phi_1^{\text{e}}(\mathbf{s}) \equiv \mathcal{V}(\delta)\phi^{\text{e}}(\mathbf{s}) \\ p(\mathbf{s},\delta) &= p(\delta)\Phi_1^{\text{o}}(\mathbf{s}) \equiv p(\delta)\phi^{\text{o}}(\mathbf{s}) \end{aligned}\ . \tag{4}$$

The third term is the elastic energy cost of un-buckling the spacer:

$$\mathcal{V}_{\delta}(\delta) \equiv \Delta(\delta)\ . \tag{5}$$

Putting everything together, we write the free energy as

$$\mathcal{V}(\mathbf{s},\delta,\mathcal{E}) = \mathcal{V}(\delta)\phi^{\text{e}}(\mathbf{s}) - \mathcal{E}p(\delta)\phi^{\text{o}}(\mathbf{s}) + \Delta(\delta)\ . \tag{6}$$

Because the sliding path **s** considered in this work is restricted to a high-symmetry path, the terms in Eq. 6 can be simplified. To describe the critical switching behavior along the one-dimensional (1D) sliding coordinate $s$, the free energy function $\mathcal{V}(s,\delta,\mathcal{E})$ can be simplified as:

$$\mathcal{V}_{\text{tot}}(s,\delta,\mathcal{E}) = V(\delta)\phi^{\text{e}}(s) - \mathcal{E}P(\delta)\phi^{\text{e}}(s) + \Delta(\delta)\ . \tag{7}$$

where the first term describes the stacking energy along the sliding path as a function of stacking $s$ and buckling $\delta$. The second term is the electrostatic coupling between the dipole moment $p(s,\delta) = P(\delta)\phi^{\mathrm{o}}(s)$ and the electric field $\mathcal{E}$. The final term $\Delta(\delta)$ represents the elastic energy, namely the energy required to deform the spacer from its equilibrium buckled configuration at $\delta = 1$. In the following, we expand the polarization only to first order $\phi^{\mathrm{o}}(s) = \Phi_1^{\mathrm{o}}(s)$. We add a small 2$^{\mathrm{nd}}$-order contribution to the stacking energy to improve the fitting to first-principles calculations: $\phi^{\mathrm{e}}(s) = \Phi_1^{\mathrm{e}}(s) - \frac{1}{8}\Phi_2^{\mathrm{e}}(s)$. As shown in of the main text, the model defined by Eq. (7) gives an excellent fit to the DFT calculations.

We impose the following realistic physical constraints on the model:

1. **Buckling Limit**: There is a critical maximum value of buckling $\delta_{\mathrm{m}}$, beyond which the ferroelectric phase becomes unstable. $\delta_{\mathrm{m}} > 1$ is defined as the the buckling for which the stacking energy surface inverts, the AB and BA stackings become unstable, and the AA stacking becomes energetically favorable, see Fig. S4. This imposes the following boundary conditions on the model parameters: $V(\delta_{\mathrm{m}}) = P(\delta_{\mathrm{m}}) = 0$. For the same reason, we also restrict $\delta$ to positive values: $\delta \geq 0$.

2. **Ideal (Flat) Limit**: In a completely flattened state ($\delta = 0$), both the effective switching barrier and the out-of-plane polarization strictly vanish: $V(0) = P(0) = 0$. This is consistent with results obtained from first-principles calculations, see Figs. S4 and S5.

3. **Equilibrium State**: We define $\delta = 1$ to correspond to the equilibrium (buckled) configuration of the spacer, which in the absence of an external electric field corresponds to an energy minimum: $\Delta'(\delta)|_{\delta=1} = 0$.

The equilibrium configuration of the system is determined by minimizing the total free energy with respect to order parameters $s$ and $\delta$, for a given value of $\mathcal{E}$. For a critical transition to occur, two mathematical criteria:

1. **Stationary Condition**: The system must reside at a local equilibrium, meaning the gradient of the energy landscape is zero:

$$\frac{\partial \mathcal{V}_{\mathrm{tot}}}{\partial s} = \frac{\partial \mathcal{V}_{\mathrm{tot}}}{\partial \delta} = 0\,. \tag{8}$$

2. **Singularity Condition**: The phase transition happens at the margin of stability, where the Hessian matrix of the state variables loses its positive definiteness:

$$\det(H) = \frac{\partial^2 \mathcal{V}_{\mathrm{tot}}}{\partial s^2} \cdot \frac{\partial^2 \mathcal{V}_{\mathrm{tot}}}{\partial \delta^2} - \left(\frac{\partial^2 \mathcal{V}_{\mathrm{tot}}}{\partial s \partial \delta}\right)^2 = 0\,. \tag{9}$$

Taking the partial derivative of $\mathcal{V}_{\mathrm{tot}}$ with respect to the sliding coordinate $s$ yields:

$$\frac{\partial \mathcal{V}_{\mathrm{tot}}}{\partial s} = -8\pi \sin(3\pi s) \cdot \left\{ V(\delta) \left[ \cos(\pi s) - \frac{3}{16}\cos(3\pi s) \right] - \mathcal{E} P(\delta) \sin(\pi s) \right\}\,, \tag{10}$$

from which we immediately see that $\sin(3\pi s) = 0 \implies s = \frac{k}{3}, k \in \mathbb{Z}$ is always a solution. $k = 1$ ($k = 2$) corresponds to the AB (BA) stacking. To evaluate the stability of this solution, we look at the second derivative evaluated at $s = k/3$:

$$\left.\frac{\partial^2 \mathcal{V}_{\mathrm{tot}}}{\partial s^2}\right|_{s=k/3} = -24\pi^2(-1)^k \left\{ V(\delta) \left[ \cos\left(\frac{k\pi}{3}\right) - \frac{3}{16}(-1)^k \right] - \mathcal{E} P(\delta) \sin\left(\frac{k\pi}{3}\right) \right\} \tag{11}$$

Now we focus on the initial energy minimum at $s = 1/3$ (AB stacking). As the field $\mathcal{E}$ increases, the system is trapped in this potential valley until the curvature vanishes and the sliding barrier disappears, described by $\frac{\partial^2 \mathcal{V}_{\mathrm{tot}}}{\partial s^2} = 0$. We note also that $\left.\frac{\partial \mathcal{V}_{\mathrm{tot}}}{\partial s}\right|_{s=1/3}$ is always 0, the mixed derivative, which appears in $\det(H)$, vanishes, and the Hessian becomes diagonal at $s = \frac{1}{3}$. Then $\det(H) = 0$ when $\frac{\partial^2 \mathcal{V}_{\mathrm{tot}}}{\partial s^2} = 0$.

From $\frac{\partial^2 \mathcal{V}_{\mathrm{tot}}}{\partial s^2} = 0$, we find that there is a critical electric field at which interlayer sliding occurs:

$$\mathcal{E}_s = \frac{11 V(\delta_s)}{8\sqrt{3} P(\delta_s)}\,. \tag{12}$$

While the system is in the $s = 1/3$ valley prior to switching, the buckling parameter $\delta$ is not fixed, but also responds to the electric field. The stationary condition for $\delta$ is:

$$\frac{\partial \mathcal{V}_{\mathrm{tot}}}{\partial \delta} = V'(\delta)\phi^{\mathrm{e}}(s) - \mathcal{E} P'(\delta)\phi^{\mathrm{o}}(s) + \Delta'(\delta) = 0\,. \tag{13}$$

At $s = \frac{1}{3}$, and the critical field $\mathcal{E}_s$, we have

$$33V(\delta)P'(\delta) + 16P(\delta)\Delta'(\delta) = 0\,. \quad (14)$$

From the ideal flat limit ($\mathcal{E} = \mathcal{E}_0$), $V(0) = P(0) = 0$, and the buckling limit ($\mathcal{E} = \mathcal{E}_\mathrm{m}$), $V(\delta_\mathrm{m}) = P(\delta_\mathrm{m}) = 0$, both $\delta = 0$ and $\delta = \delta_\mathrm{m}$ trivially satisfy this equation. We denote another possible solution to Eq. 13 as $\delta_s \in (0, \delta_\mathrm{m})$, which corresponds to the onset of buckling accompanied by interlayer sliding, with associated critical field $\mathcal{E}_s$. In this case, sliding occurs before the spacer is fully flattened.

Depending on the ordering of $\mathcal{E}_\delta^0$, $\mathcal{E}_\delta^\mathrm{m}$, and $\mathcal{E}_s$, different types of switching behavior occur, as described in the main text and the following section.

The superlubric ferroelectric system has three critical values of $\delta$: $\delta_0 = 0$ (the completely flattened state), $\delta_\mathrm{m}$ (the saturated buckled state), and potentially $\delta_s$ (the critical buckling value right before switching). Depending on the relative order of the critical electric fields associated with these values of $\delta$, there are four distinct types of ferroelectric hysteresis loops that can occur, as described in the main text.

## IV. FITTING THE MODEL TO FIRST-PRINCIPLES CALCULATIONS

Fig. S4 shows how the energy as a function of sliding depends on the buckling parameter $\delta$ in MBM. Going from the equilibrium state ($\delta = 1$) to an ideal flat spacer ($\delta = 0$), the energy surface softens and eventually becomes completely flat, see Fig. S4 (a). The energy surface also softens upon enhancing the buckling beyond its equilibrium value, and eventually inverts, see Fig. S4 (b). For MBM, the ferroelectric phase becomes unstable at $\delta_\mathrm{m} \approx 1.5$.

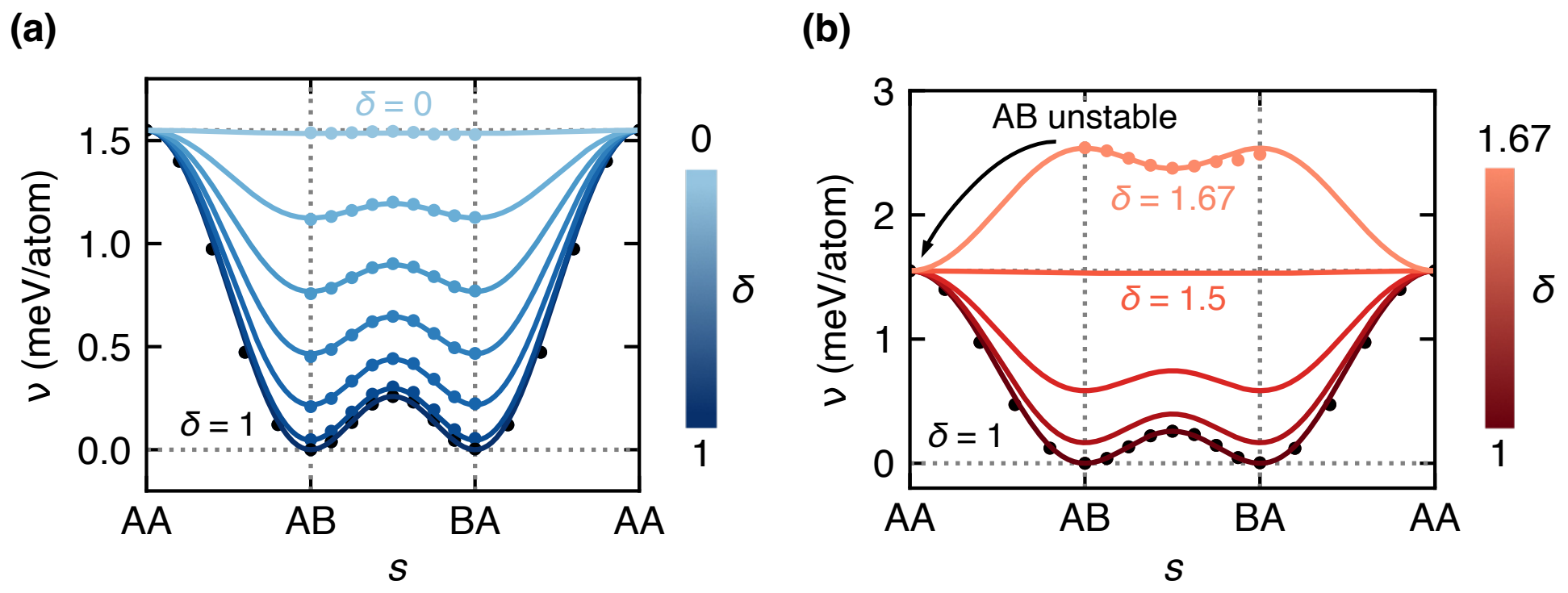


FIG. S4. **Sliding energy surface as a function of $\delta$.** **(a)** Relative energy as a function of sliding for different fixed values of $\delta$ between 0 and 1. The data shows results obtained from first-principles calculations, and the solid lines show the corresponding fit of the model. Calculations were performed between the AB and BA stackings, except for $\delta = 1$, where the entire unit cell diagonal was parametrized. The model yields a good fit to the first-principles calculations along the entire unit cell diagonal. **(b)** Relative energy as a function of sliding for different fixed values of $\delta$ for $\delta > 1$. The ferroelectric phase becomes unstable for $\delta_\mathrm{m} \approx 1.5$ (as shown by the black arrow), beyond which the energy surface becomes inverted.

Figure S5 shows the functional dependence of $V(\delta)$, $\Delta(\delta)$, and $P(\delta)$ on $\delta$, as obtained from first-principles calculations. The lines in the figure represent the fitting curves obtained from polynomial fits truncated at the cubic order, with the fitted parameters listed in Table. I. Notably, as shown in Fig. S5 (a), $V(\delta)$ approaches zero at $\delta = 1.5$, corresponding to the nearly flat potential energy surface in Fig. S4 (b). The buckling exceeding $\delta = 1$ at which $V(\delta) = 0$ is defined as the maximum buckling $\delta_\mathrm{m}$, beyond which the ferroelectric stacking states of SL-sFEs can no longer be stabilized. For $\delta > \delta_m$, the system slides into a non-ferroelectric stacking state, accompanied by the disappearance of the polarization $P(\delta)$, as shown in Fig. S5 (c).

TABLE I. Polynomial fitting results for $V(\delta)$, $D(\delta)$, and $P(\delta)$, where the fitting function is $f(\delta) = a_0 + a_1\delta + a_2\delta^2 + a_3\delta^3$.

| fitting parameters | $a_0$ | $a_1$ | $a_2$ | $a_3$ |
|---|---|---|---|---|
| $V(\delta)$ | 0.026 | 0.2117 | 0.393 | -0.3567 |
| $\Delta(\delta)$ | 1.4906 | -1.9535 | -0.7249 | 1.1565 |
| $P(\delta)$ | -0.0027 | -0.4126 | -0.0202 | -0.0736 |

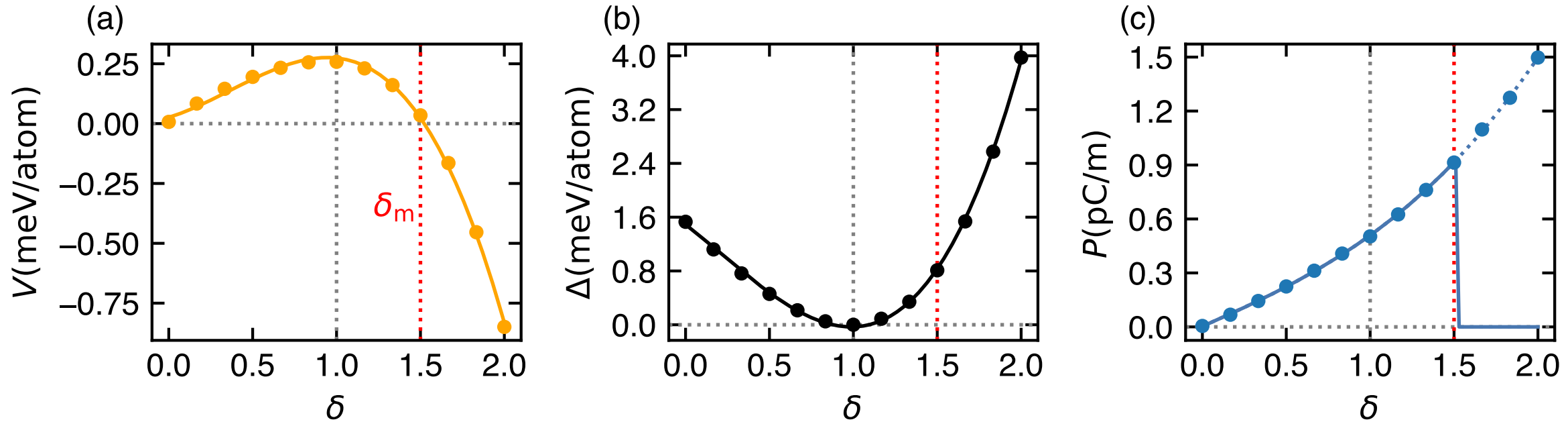


FIG. S5. **Fitting results of the model parameters.** (a–c) correspond to the three $\delta$-dependent variables $V(\delta)$, $\Delta(\delta)$, and $P(\delta)$ in Eq. (1), respectively. The red dashed line indicates the maximum buckling $\delta_m$ that maintains a stable ferroelectric stacking. When $\delta$ exceeds $\delta_m$, the energy of the ferroelectric stacking becomes unstable. In this regime, $V(\delta)$ in (a) becomes negative and $P(\delta)$ becomes zero.



\end{document}